\documentclass[preprint,prc,aps,epsfig]{revtex4}
\usepackage{psfig}

\def\beq{\begin{equation}}
\def\eeq{\end{equation}}
\def\beqa{\begin{eqnarray}}
\def\eeqa{\end{eqnarray}}
\def\MeV{\nobreak\,\mbox{MeV}}
\def\GeV{\nobreak\,\mbox{GeV}}

\def\pli{p^\prime}
\def\mli{{M^\prime}^2}
\def\mmi{{M^\prime}}
\def\al{\alpha}

\begin{document}

\title{$J/\psi$ couplings to open charm mesons from QCD sum rules}

\author{R.D. Matheus, F.S. Navarra, M. Nielsen and R. Rodrigues da Silva}
\affiliation{Instituto de F\'{\i}sica, Universidade de S\~{a}o Paulo\\
 C.P. 66318,  05315-970 S\~{a}o Paulo, SP, Brazil}

\begin{abstract}
We employ  QCD sum rules to calculate the $J/\psi$ form factors and 
coupling constants with charmed mesons,  by studying three-point  correlation
functions. In particular we consider the $J/\psi D^*D$ and $J/\psi DD$
vertices.
 We determine the momentum dependence of the form factors for kinematical 
conditions where $J/\psi,~D$ or $D^*$ are off-shell. Extrapolating to the
 pole of each one of the so obtained form factors, we determine the coupling
constants. For  both couplings our results (within the errors) 
are compatible with estimates based on constituent quark model.
\end{abstract}

\pacs{PACS: 12.38.Lg, 14.40.Lb, 13.75.Lb}

\maketitle

\section{Introduction}

Since their begining in 1998, effective Lagrangian theories for charmed mesons
were constructed to give us a better understanding of the interaction
between $J/\psi$ and light mesons in nuclear matter. Although the original
motivation is still very strong, specially in view of the vigorous ongoing
RHIC program, this new sub-area of hadron physics introduced new questions
which are now in debate. Some of them are: What are the coupling constants
involving heavy mesons? Which are the form factors appearing in these vertices?
To what extent are we allowed to use SU(4) symmetry?
Moreover, interactions of charmed mesons offer a testing ground for the 
predictions of heavy quark effective theory (HQET) \cite{hqet}. As pointed 
out in ref.~\cite{dea}, a consequence of the spin symmetry of HQET is
that
\beq
g_{\psi DD^*}={g_{\psi DD}\over m_D}.
\label{regg}
\eeq
We will in this work calculate both couplings independently and will have 
the opportunity to check the relation above.

Charmed hadrons are composites of the underlying quarks whose effective fields
describe point-like physics only when all the interacting particles are
on mass-shell. When at least one of the particles in a vertex is off-shell,
the finite size effects of  hadrons become important. Therefore, the
knowledge of  form factors in hadronic vertices is of crucial importance
to estimate any amplitude using hadronic degrees of freedom. 
This work is devoted to the study of the $J/\psi D^*D$ form factor, which
is important, for instance, in the evaluation of the  dissociation cross 
section of $J/\psi$ by pions and $\rho$ mesons using effective 
Lagrangians \cite{osl,nnr,haga2}.

The $J/\psi D^*D$  coupling has been studied by some authors 
using different approaches: vector meson dominance model plus relativistic
potential model \cite{osl} and constituent  quark models \cite{dea}.
Unfortunately, the numerical results from these 
calculations may differ by almost a factor two. The relevance of this 
difference can not be underestimated since the cross sections are proportional
to the square of the coupling constants. In ref.~\cite{haga2} it was shown
that the use of different coupling constants and form factors can lead to 
cross
sections that differ by more than one order of magnitude, and that can even 
have a different behavior as a function of $\sqrt{s}$. 

In previous works we have used the QCD sum rules (QCDSR) to
study the $D^*D\pi$ \cite{nos1, nos2}, $DD\rho$ 
\cite{nos4} and $J/\psi DD$ \cite{nos3} form factors, considering two
different mesons off mass-shell. In these works the QCDSR results for 
the form factors were parametrized by analytical forms
such that the respective extrapolations to the off-shell meson poles
provided consistent values for the corresponding coupling constant.
In this work we use the QCDSR approach to evaluate the 
$J/\psi D^*D$  form factors and extend the  procedure described above to
estimate the $J/\psi D^*D$ coupling constant considering each one of these 
mesons off-shell, calculating the associated form factor, and then 
extrapolating the three obtained form factors to the corresponding
on-shell masses. We will apply this
method to both $J/\psi DD$ and $J/\psi D^*D$ vertices.

\section{The QCD sum rule approach}

The three-point function associated with the hadronic vertex $J/\psi DD^*$, 
where $J/\psi$ and
$D$ are the incoming and outgoing external mesons with momentum $p$ and 
$\pli$ respectively and $D^*$
is the off-shell meson, with momentum $q=\pli-p$ is given by
\beqa
\Gamma_{\mu\nu}(p,\pli)=\int d^4x \, d^4y \,   
\, e^{ip^\prime.x} \, e^{-iq.y} 
\langle 0|T\{j_{D}(x)
{j_\nu^{D^*}}^\dagger(y){j_\mu^{\psi}}^\dagger(0)\}|0\rangle\;,
\label{cor}
\eeqa
where the interpolating fields are  $j_\mu^{\psi}=
\bar{c}\gamma_\mu c$, $j_\nu^{D^*}=\bar{q}\gamma_\nu c$ and
$j_D=i\bar{q}\gamma_5 c$  with $q$ and $c$ being a light quark and 
the charm quark fields.

The fundamental assumption of the QCD sum rule approach \cite{svz,rry}
is the principle of duality, {\it i.e.}, it is assumed that there is an 
interval over which a correlation function
may be equivalently described at the quark level and at
the hadron level. Therefore, the procedure of the QCD sum rule
technique is the following: on one hand we calculate the correlator 
in Eq.(\ref{cor}) at the quark level in terms of quark and gluon fields. 
On the
other hand, the correlator is calculated at the hadronic level
introducing hadron characteristics, giving sum rules
from which a hadronic quantity can be estimated.

The hadronic side of the correlation function, 
$\Gamma_{\mu\nu}(p,p^\prime)$,
is obtained by the consideration of $J/\psi$, $D^*$ and $D$ state 
contributions to
the matrix element in Eq.~(\ref{cor}):
\beqa
\Gamma_{\mu\nu}^{(phen)}(p,\pli)={m_D^2f_D\over m_c}m_{D^*}f_{D^*}m_{\psi}
f_{\psi}{g_{\psi DD*}^{(D^*)}(q^2)
\epsilon_{\alpha\beta
\mu\nu}p^\alpha{\pli}^\beta\over (q^2-m_{D^*}^2)(p^2-m_\psi^2)({\pli}^2
-m_D^2)}+ \mbox{h.~r.}\; ,
\label{phen}
\eeqa
where h.~r. means higher resonances.

The meson decay constants appearing in Eq.~(\ref{phen}) are defined
by the vacuum to meson transition amplitudes:
\beq
\langle 0|j^{(D)}|D\rangle={m_D^2f_{D}\over m_c}\;,
\label{fd}
\eeq
and
\beq
\langle V(p,\epsilon)|j^\dagger_\alpha|0\rangle=m_{V}f_{V}\epsilon^*_\alpha
\; ,
\label{fv}
\eeq
for the vector mesons $V=J/\psi$ and $V=D^*$. The form factor we want
to estimate is defined through the
vertex function for an off-shell $D^*$ meson:
\beq
\langle \psi(p,\lambda)|D(\pli)D^*(q,\lambda^\prime)\rangle
=g_{\psi DD*}^{(D^*)}(q^2)
\epsilon^{\alpha\beta\gamma\delta}\epsilon_\alpha^\lambda(p)
\epsilon_\gamma^{\lambda^\prime}(q)\pli_{\beta}q_{\delta}\;.
\label{ver}
\eeq

The contribution of higher resonances and continuum in Eq.~(\ref{phen})
will be taken into account as usual in the standard form of 
ref.~\cite{io2}.

The QCD side, or theoretical side, of the correlation function is evaluated
 by
performing Wilson's operator product expansion (OPE) of the operator
in Eq.~(\ref{cor}). 
Writing $\Gamma_{\mu\nu}$ in terms of the invariant
amplitude:
\beq
\Gamma_{\mu\nu}(p,\pli)=\Lambda(p^2,{\pli}^2,q^2)\epsilon_{\alpha\beta
\mu\nu}p^\alpha{\pli}^\beta\;,
\eeq
we can write a double dispersion relation for $\Lambda$,
over the virtualities $p^2$ and ${\pli}^2$
holding $q^2$ fixed:
\beq
\Lambda(p^2,{\pli}^2,q^2)=-{1\over4\pi^2}\int ds
du~ {\rho(s,u,q^2)\over(s-p^2)(u-{\pli}^2)}\;,
\label{dis}
\eeq
where $\rho(s,u,q^2)$ equals the double discontinuity of the amplitude
$\Lambda(p^2,{\pli}^2,q^2)$ on the cuts $4m_c^2\leq s\leq\infty$,
$m_c^2\leq u\leq\infty$,
We consider diagrams up to dimension three which include the perturbative 
diagram and the quark condensate.

To improve the matching between the two sides 
of the sum rules, we perform a double Borel transformation  in both 
variables
$P^2=-p^2\rightarrow M^2$ and ${P^\prime}^2=-{\pli}^2\rightarrow\mli$. 
We get the following sum rule:
\beqa
C~\frac{g_{\psi DD^{*}}^{(D^*)}(t)}{(t- m_{D^*}^2)}
e^{-\frac{m_{D}^2}{\mli}}
e^{-\frac{m_{\psi}^2}{M^2}}
&=&\frac{1}{4\pi^2}
\int_{4m^2}^{s_0}
\int_{u_{min}}^{u_0}
dsdu~
\rho^{(D^*)}(u,s,t)e^{-\frac{s}{M^{2}}}e^{-\frac{u}{\mli}}\Theta(u_{max}-u)
\nonumber\\
&+&{m_c\langle g_s^2 G^2\rangle\over2\pi^2}\Pi_G(t,M^2,\mli),
\label{ffd}
\eeqa
where $C={m_D^2f_D\over m_c}m_{D^*}f_{D^*}m_{\psi}
f_{\psi}$, $t=q^2$,
\beq
\rho^{(D^*)}(u,s,t)=
\frac{3m_c}{\sqrt{\lambda}}
\left(
1+\frac{s\lambda_2}{\lambda}
\right),
\eeq
with $\lambda =(u+s-t)^2 -4us$, $\lambda_2 =u+t-s +2m_c^2$ and
\beq
u^{max}_{min}=\frac{1}{2m_c^2}
\left[
-st+ m_c^2(s+ 2t) 
\pm 
\sqrt{s(s-4m_c^2)(t-m_c^2)^2}
\right].
\eeq
The last term in Eq.~(\ref{ffd}) gives the gluon condensate contribution. 
The full
expression for the gluon condensate contribution is given in Appendix A.

In Eq.~(\ref{ffd}) we have 
transferred to the QCD side the higher resonances contributions 
through the introduction of
the continuum thresholds $s_0$ and $u_0$. The sum rules for the form factors
considering the mesons $J/\psi$ and $D$ as off-shell mesons can be obtained
in a similar way. In the case that the $D$ meson is off-shell we get
basically the same sum rule, the only difference being the exchange between
$m_D^2$ and $m_{D^*}^2$ in the left hand side of Eq.~(\ref{ffd}). The 
spectral density is exactly the same \cite{bjp}. In the case that $J/\psi$
is off-shell we get
\beq
C~\frac{g_{\psi DD^{*}}^{(J/\psi)}(t)}{(t- m_{\psi}^2)}
e^{-\frac{m_{D}^2}\mli}
e^{-\frac{m_{D^{*}}^2}{M^2}}
=\frac{1}{4\pi^2}
\int_{m^2}^{s_0}
\int_{u_{min}}^{u_0}
dsdu
\rho^{(J/\psi)}(u,s,t)e^{-\frac{s}{M^{2}}}e^{-\frac{u}{\mli}}\Theta(u_{max}
-u),
\label{ffpsi}
\eeq
with 
\beqa
\rho^{(J/\psi)}(u,s,t)=
\frac{3m_c}{\lambda^{3/2}}
\left[
(u-s)^2 -t(u+ s- 2m_c^2)
\right]-4\pi^2<\bar{q}q>\delta(s-m_c^2)\delta(u-m_c^2),
\eeqa
and
\beq
u^{max}_{min}=\frac{1}{2m_c^2}
\left[
-st+ m_c^2(2s+ t) 
\pm 
\sqrt{t(t-4m_c^2)(s-m_c^2)^2}
\right].
\eeq

\section{Form factors in the $J/\psi$ vertices}
\subsection{$J/\psi DD^*$}

The parameter values used in all calculations are: $m_D=1.87\,\GeV$, 
$m_{D^*}=2.01\,\GeV$, $m_\psi=3.1\,\GeV$, $f_{J/\psi}=(405 \pm 15)\MeV $
\cite{PDG}. For $f_D$ and $f_{D^*}$ we use the values evaluated in the
two-point QCD sum rules under the same kind of approximations \cite{bbkr}:
$f_{D}=(170 \pm 10)~\MeV $, $f_{D^{*}}=(240 \pm 20)~\MeV $. For the charm 
quark mass, quark condensate and gluon condensate we use the values normally 
used in QCD sum rules calculations \cite{svz,rry}
$m_c=1.3\,\GeV$, $\langle\overline{q}q\rangle\,=\,-(0.23)^3\,\GeV^3$,
 $\langle g_s^2 G^2\rangle\,=\,0.5\,\GeV^4$ 
For the continuum thresholds we take $s_0=(m_\psi+\Delta_s)^2$ 
and $u_0=(m_{D(D^*)}+\Delta_u)^2$ for the sum rules when $D(D^*)$ is 
off-shell, and $s_0=(m_{D^*}+\Delta_s)^2$ 
and $u_0=(m_{D}+\Delta_u)^2$ for the sum rule when $J/\psi$ is off-shell.
We take $\Delta_s=\Delta_u=(0.5\pm0.1)\GeV$.

\begin{figure}[htb]
\centerline{\psfig{figure=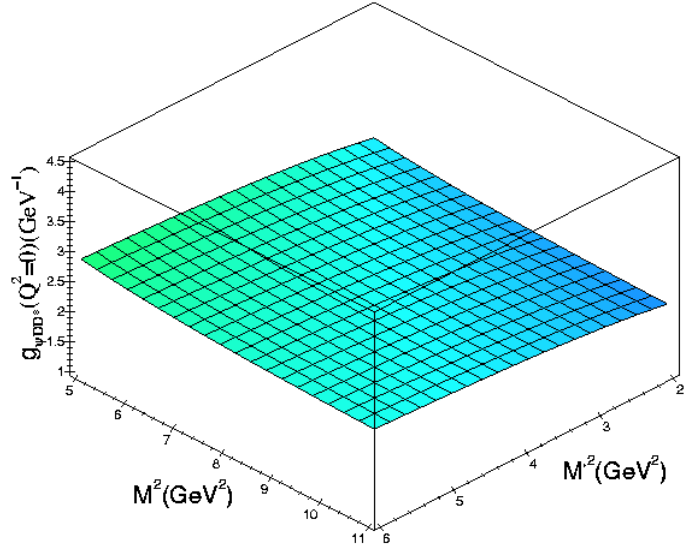,width=9cm,angle=0}}
\protect\caption{$M^2$ and $\mli$ dependence of 
$g_{\psi DD^*}^{(D^*)}(Q^2=0)$.}
\label{fig1}
\end{figure}

We first discuss the $J/\psi DD^*$ form factor with an off-shell $D^*$ 
meson. 
Fixing $Q^2=-q^2=0$ and $\Delta_s=\Delta_u=0.5~\GeV$
we show in Fig.~1 the Borel dependence of the form 
factor
$g_{\psi DD^*}^{(D^*)}(0)$. We see that we get a very good stability for 
the form factor as a function of the two
independent Borel parameters in the considered Borel regions.
\begin{figure}[h]
\centerline{\psfig{figure=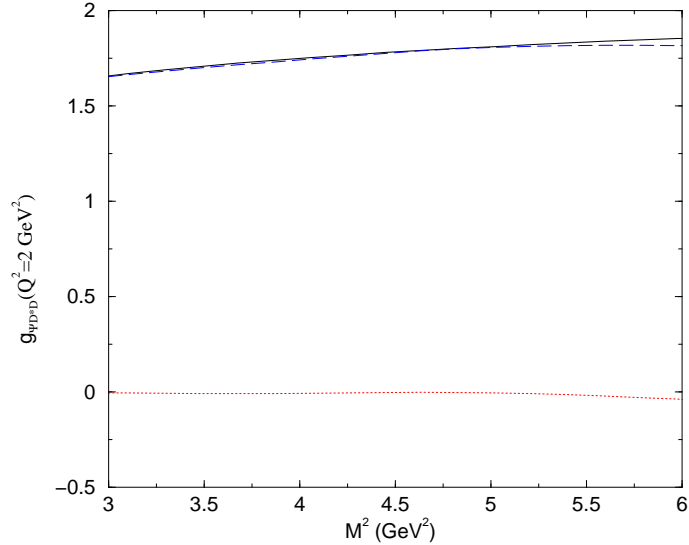,width=9cm,angle=0}}
\protect\caption{$\mli$ dependence of perturbative contribution (dashed line)
and the gluon condensate contribution (dotted line) to 
$g_{J/\psi DD^*}^{(D^*)}$ at $Q^2=2.0\,\GeV^2$. The solid line gives the
final reslt for the form factor.}
\label{fig2}
\end{figure}
The same kind of stability is obtained for 
other values of $Q^2$ and for the other two form factors \cite{bjp}.

In Fig.~2 we show the perturbative (dashed line) and the gluon condensate
(dotted line) contributions to the form factor $g_{J/\psi DD^*}^{(D^*)}(Q^2)$ 
at $Q^2=2~\GeV^2$ as a function of the Borel mass $\mli$ at a fixed ratio 
$\mli/M^2={m_{D}^2/m_\psi^2}$.
We see that the gluon condensate contribution is negligible, as compared
with the perturbative contribution. The same kind of behaviour is obtained for 
other values of $Q^2$. This is a very interesting result since it might be
indicating a convergence of the OPE.

In Fig.~3 we show the behavior of the form factors 
$g_{J/\psi DD^*}(Q^2)$ at $Q^2=2~\GeV^2$ 
as a function of the Borel mass $\mli$. 
The solid line gives 
$g_{J/\psi DD^*}^{(D^*)}$ at a fixed ratio $\mli/M^2={m_{D}^2
/m_\psi^2}$. The dashed line gives
$g_{J/\psi DD^*}^{(D)}$ at a fixed ratio $\mli/M^2={m_{D^*}^2/
m_\psi^2}$, and the dotted line gives $g_{J/\psi DD^*}^{(J/\psi)}$ 
at a fixed ratio $\mli/M^2={m_{D}^2/ m_{D^*}^2}$.

\begin{figure}[h]
\centerline{\psfig{figure=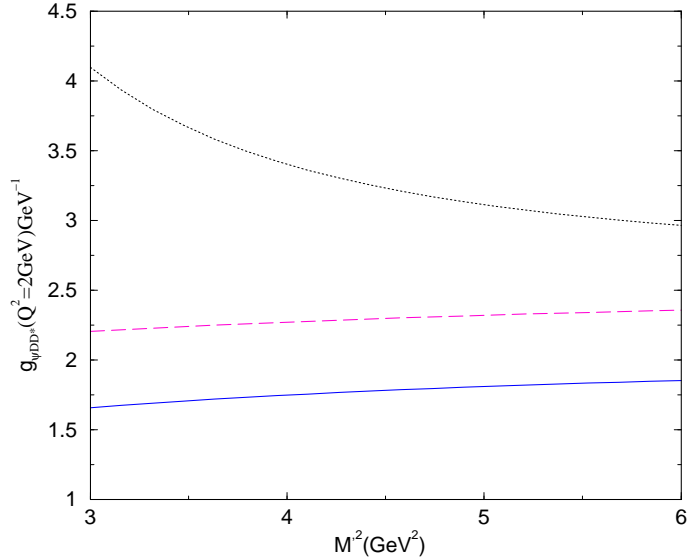,width=9cm,angle=0}}
\protect\caption{$\mli$ dependence of $g_{J/\psi DD^*}^{(D^*)}$ (solid 
line), $g_{J/\psi DD^*}^{(D)}$ (dashed line) and  $g_{J/\psi DD^*}^{(J/
\psi)}$ 
(dotted line) for $Q^2=2.0\,\GeV^2$.}
\label{fig3}
\end{figure}
We can see that the QCDSR results for $g_{J/\psi DD^*}^{(D^*)}$ and
$g_{J/\psi DD^*}^{(D)}$ are very stable
in the interval $3\leq \mli\leq6\,\GeV^2$. In the case of 
$g_{J/\psi DD^*}^{(J/\psi)}$ the stability is not as good as for the other
form factors, but it is still acceptable.

\begin{figure}[htb]
\centerline{\psfig{figure=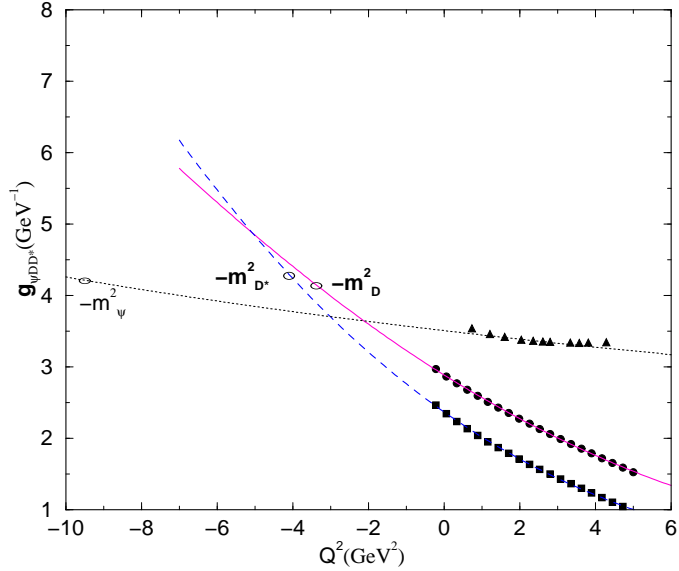,width=9cm,angle=0}}
\protect\caption{Momentum dependence of the $J/\psi DD^*$ form factors. The 
dotted, dashed and solid lines give the 
parametrization of the QCDSR results (triangles, squares and circles) 
through Eqs.~(\protect\ref{papsi}),
(\protect\ref{pads}) and (\protect\ref{pad}) respectively.}
\label{fig4}
\end{figure}

Fixing $M^2$ and $\mli$ at the values of the incoming and outgoing
meson masses we  show, 
in Fig.~4, the momentum dependence of the QCDSR results for the form 
factors 
$g_{\psi DD^*}^{(D)}$, $g_{\psi DD^*}^{(D^*)}$ and $g_{\psi DD^*}^{
(J/\psi)}$ through the circles, squares and triangles respectively.
Since the present approach 
cannot be used at $Q^2<<0$, in order to extract the $g_{\psi DD^*}$
coupling from the form factors we need to extrapolate the curves to 
the mass of the off-shell meson, shown as open circles in Fig.~4.

In order to do this extrapolation we fit 
the QCDSR results  with an analytical expression. We tried to fit
our results to a monopole form, since this is very often used 
for form factors, but the fit was only good for $g_{\psi DD^*}^{(J/\psi)}$. 
For $g_{\psi DD^*}^{(D)}$ and $g_{\psi DD^*}^{(D^*)}$
we obtained  good fits using  a Gaussian form. We get:
\beq
g^{(J/\psi)}_{\psi DD^{*}}(Q^2)=
\frac{200}{57+Q^{2}}=4.2{57-m_\psi^2\over 57+Q^2},
\label{papsi}
\eeq
\beq
g^{(D^{*})}_{\psi DD^{*}}(Q^2)=
20~
e^{-\frac{(Q^2+27)^{2}}{345}},
\label{pads}
\eeq
\beq
g^{(D)}_{\psi DD^{*}}(Q^2)=
13~
e^{-\frac{(Q^2+26)^{2}}{450}}.
\label{pad}
\eeq
These fits are also shown in Fig.~4 through the dotted, dashed and solid 
lines respectively. From Fig.~4 we see that all three form factors lead
to compatible values for the coupling constant when the form factors are 
extrapolated to the off-shell meson masses (open circles in Fig.~4).
 
All results showed above were obtained using $\Delta_s=\Delta_u=0.5~\GeV$.
In Fig.~5 we use the form factor $g^{(D)}_{\psi DD^{*}}(Q^2)$ to ilustrate
the dependence of our results with the continuum thresholds.

\begin{figure}[htb]
\centerline{\psfig{figure=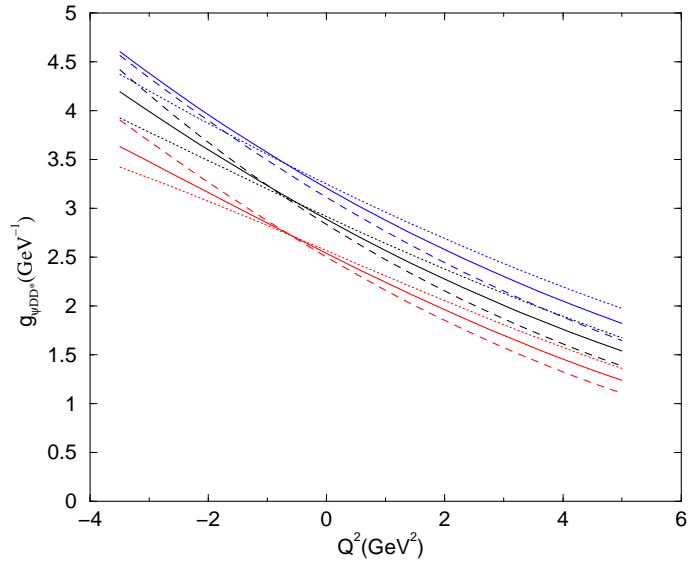,width=9cm,angle=0}}
\protect\caption{Continuum threshold dependence of the form factor
$g^{(D)}_{\psi DD^{*}}(Q^2)$. The dashed solid and dotted lines give the 
parametrization of the QCDSR results for  $\Delta_s=0.4~\GeV,~0.5~\GeV$
and $0.6~\GeV$ respectively. The lower, intermediate and upper set of curves 
show the results for $\Delta_u=0.4~\GeV,~0.5~\GeV$
and $0.6~\GeV$ respectively.}
\label{fig5}
\end{figure}

In Fig.~5 there are three sets of three curves. They show the extrapolation
of the QCD sum rule results, through  parametrizations similar to the
one given in Eq.~(\ref{pad}), to all kinematical region. The lower, 
intermediate and upper sets were obtained using $\Delta_u=0.4~\GeV,~0.5~\GeV$
and $0.6~\GeV$ respectively. The  dashed solid and dotted lines 
in each set were obtained using $\Delta_s=0.4~\GeV,~0.5~\GeV$
and $0.6~\GeV$ respectively. We see that the dispersion in the region
$0\leq Q^2\leq4.5~\GeV^2$, where we have the QCDSR points, does not
lead to a bigger dispersion at $Q^2=-m_D^2$, where the coupling constant
is extracted. Therefore, the extrapolation procedure used here does not
act as a lever effect for the uncertainties.

Considering the uncertainties in  
the continuum threshold  and the difference 
in the values of the coupling constants extracted when the $D$, $D^*$ or 
$J/\psi$ mesons
are off-shell, our result for the $J/\psi DD^*$ coupling constant is:

\beq
g_{\psi DD^{*}}=(4.0 \pm 0.6)\mbox{GeV}^{-1}.
\label{fi}
\eeq

From the parametrizations in Eqs.~(\ref{papsi}),
(\ref{pads}) and (\ref{pad}) we can also extract the cutoff parameter,
$\Lambda$, associated with the form factors. The general expression for
the Gaussian parametrization is: $A\exp{[-(Q^2+B)^2/\Lambda^4]}$, which 
gives $\Lambda\sim 4.5~\GeV$ when the off-shell meson is $D$ or $D^*$. For
the monopole parametrization the general expression is: $g[(\Lambda^2-m^2)/(
\Lambda^2+Q^2)]$. Therefore, for an off-shell $J/\psi$ we get
$\Lambda\sim7.5~\GeV$. It is very interesting to notice that the values
of the cutoffs obtained here follow the same trend as observed in 
refs.~\cite{nos2,nos4,nos3}: the value of the 
cutoff is  directly associated with the mass of the off-shell meson probing
the vertex. The form factor is harder if the off-shell meson is heavier,
implying that the size of the vertex depends on the exchanged meson.
Therefore, a heavy meson (like $J/\psi$) will see the vertex 
approximately as point like,
whereas a light meson will see its extension.

In Table I we show the results obtained for the same coupling constant 
using different approaches.
While our result is compatible with the coupling obtained using the
constituent  quark meson model \cite{dea}, it is half of the value
obtained with the vector meson dominance (VMD) model  plus relativistic
potential model \cite{osl}. It is important to mention that in the case
of the VDM model, the coupling is extracted for an off-shell
$J/\psi$ with $Q^2=0$. However, since our $g_{\psi DD^{*}}^{(J/\psi)}(Q^2)$
form factor depends weakly on $Q^2$, the value extracted at
$Q^2=0$ is still very close to the value in Eq.~(\ref{fi}).

\subsection{$J/\psi DD$}

In ref.~\cite{nos3} we have studied the $J/\psi DD$ vertex, and we have
calculated the form factors considering two cases:
i) one $D$ and ii) the $J/\psi$ as off-shell mesons. We obtained
\beq
g^{(J/\psi)}_{\psi DD}(Q^2)=
\frac{306}{63+Q^{2}}=5.7{63-m_\psi^2\over 63+Q^2},
\label{paj}
\eeq
\beq
g^{(D)}_{\psi DD}(Q^2)=
15~
e^{-\frac{(Q^2+20)^{2}}{250}},
\label{padd}
\eeq
which lead to the coupling
\beq
g_{\psi DD}=5.8\pm0.9\,.
\label{fi1dd}
\eeq

The numerical values presented here are a little different from the values
in ref.~\cite{nos3}, due to the fact that there we made a mistake
in the value of $f_{J/\psi}$, and have used
$f_{J/\psi}=270~\MeV$ in ref.~\cite{nos3} .

\section{Conclusions}

We have used the method of QCD sum rules  to compute form 
factors and coupling constants in the $J/\psi D D^*$ and $J/\psi DD$
vertices. We have first analyzed the $J/\psi D D^*$ vertex and have 
considered three cases: i) off-shell $D$, ii) off-shell $D^*$
and iii) off-shell $J/\psi$. In the three cases we have fitted
the QCDSR results with analitycal forms and extracted the coupling constant.
Our results for the coupling show once more 
that this method is robust, yielding numbers which are approximately the 
same regardless of which particle we choose to be off-shell and depending 
weakly on the choice of the continuum threshold. As for the form factors,
we obtain a harder form factor when the off-shell particle is 
heavier. The same comments can be made for the $J/\psi DD$ vertex.

In this work we have not considered  $\alpha_s$ corrections, which might be
not negligible in the case of heavy quarks. However, as shown for instance
in ref.~\cite{khod}, in the calculation
of $B^*B\pi$ and $D^*D\pi$ coupling constants, $\alpha_s$ corrections are
more important in the beauty case than in the charm case. In their caculation
the inclusion of $\alpha_s$ corrections have changed $g_{D^*D\pi}$ by
about 10\%, which is smaler than the theoretical uncertainties. Therefore,
we postpone the evaluation of the $\alpha_s$ corrections to a future work.

As a closing remark we come back to Eq.~({\ref{fi1dd}) from where we get 
$g_{\psi DD}/m_D=(3.1\pm0.5)~\GeV^{-1}$,
which is in agreement (considering the uncertainties) with our result
for $g_{\psi DD^*}$ in Eq.~(\ref{fi}). Therefore, our QCDSR results for the
couplings $g_{\psi DD}$ and $g_{\psi DD^*}$ obey the HQET relation
given in Eq.~(\ref{regg}).

In Table I we present our final results and compare them with other
calculations.
\begin{center}
\begin{tabular}{|c|c|c|c|}
\hline
coupling & this work & ref.~\cite{osl} & ref.~\cite{dea}\\
\hline
 $g_{\psi DD^*}$ (GeV)$^{-1}$ & 4.0 $\pm$ 0.6 & 8.0 $\pm$ 0.6& 4.05 $
\pm$ 0.25\\
 $g_{\psi DD}$  & 5.8 $\pm$ 0.8 & 7.7& 8.0 $\pm$ 0.5\\
\hline 
\end{tabular}
\end{center}
\begin{center}
{\small{\bf TABLE I:}  Values of the coupling constants
evaluated using different approaches.}
\end{center}

We see that while our result for $g_{\psi DD^*}$
is compatible with the coupling obtained using the
constituent  quark meson model \cite{dea}, this is not the case
for $g_{\psi DD}$. The values of the couplings
obtained with the vector meson dominance model  plus relativistic
potential model \cite{osl} are bigger than ours for both couplings.
The origin of these discrepancies deserves further investigation.

\section*{Acknowledgments}

This work was supported by CNPq and FAPESP.

\appendix\section{Gluon condensate contribution}

\begin{figure}[htb]
\centerline{\psfig{figure=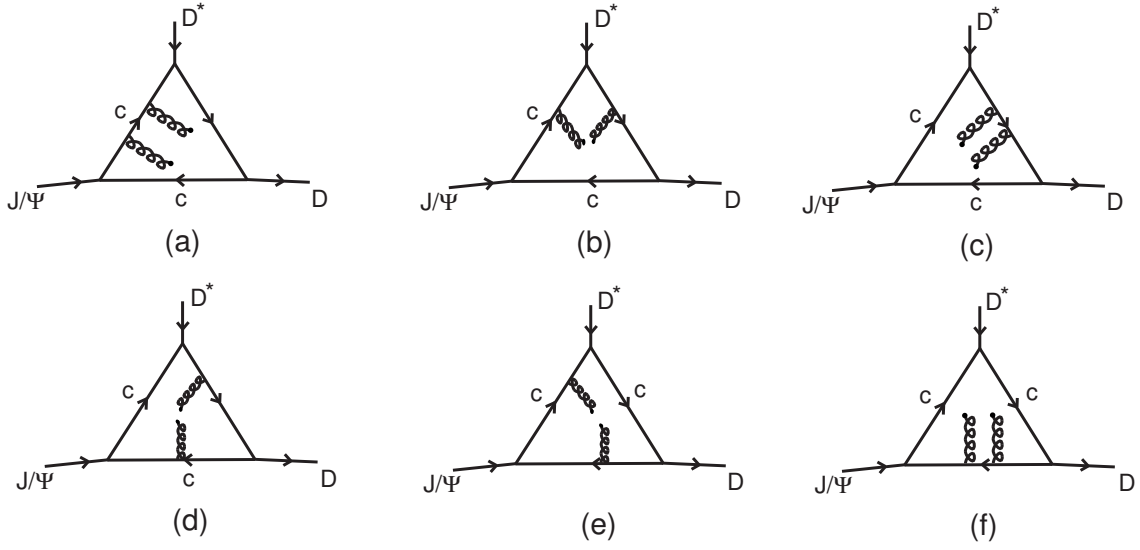,width=15cm,angle=0}}
\protect\caption{Gluon condensate diagrams contributing to the correlation
function in Eq.~(\ref{cor}).}
\label{fig9}
\end{figure}
In Fig. 8 we show the gluon condensate diagrams contributing to the
correlation function in Eq.~(\ref{cor}). We give bellow the expressions
for all diagrams contribution in terms of $\Pi_G(t,M^2,\mli)$ appearing
in Eq.~(\ref{ffd}).
\beqa
&&\Pi_G^{(a)}(t,M^2,\mli)=-{{\mmi}^6\over M^2(M^2+\mli)^2}\int_{1/\mli}^\infty
~d\al ~e^{-{m_c^2\al(M^2+\al{\mmi}^4)\over M^2(\al\mli-1)}}
~e^{t(\al\mli-1)\over M^2+\mli}\bigg[{1\over2}+
\nonumber\\
&-&\left.{m_c^2\over6}{\al(M^2+\al{\mmi}^4)\over M^2(\al\mli-1)}\right],
\eeqa

\beqa
&&\Pi_G^{(b)}(t,M^2,\mli)=-{\mli\over6 M^2(M^2+\mli)}\int_{1/\mli}^\infty
~d\al ~e^{-{m_c^2\al(M^2+\al{\mmi}^4)\over M^2(\al\mli-1)}}
~e^{t(\al\mli-1)\over M^2+\mli}\bigg[\al t\mli+
\nonumber\\
&+&{1\over\al\mli-1}\bigg(-\mli+2\al{\mmi}^4-M^2(3-5\al\mli+\al^2{\mmi}^4)
-m_c^2\al(M^2+\mli(2\al\mli+
\nonumber\\
&-&1)+\al\mli\left({m_c^2\al^2\mli\over M^2}{(M^2(M^2+2\mli)-\al{\mmi}^6(
\al\mli-2))\over(\al\mli-1)^2}+M^2+3\al{\mmi}^4+\right.
\nonumber\\
&+&\left.\left.\left.
{t\mli(\al M^2+1)(M^2+\al{\mmi}^4)\over(M^2+\mli)^2}+{\mli(M^2+\al{\mmi}^4)
(2\al\mli+\al M^2-1))\over(M^2+\mli)(\al\mli-1)}\right)\right)\right],
\eeqa

\beq
\Pi_G^{(c)}(t,M^2,\mli)=0,\;\;\;\mbox{ since it is proportional to }m_q,
\eeq

\beqa
&&\Pi_G^{(d1)}(t,M^2,\mli)=-{\mli\over6 M^2}\int_{1/\mli}^\infty
~d\al ~e^{-{m_c^2\al(M^2+\al{\mmi}^4)\over M^2(\al\mli-1)}}
~e^{t(\al\mli-1)\over M^2+\mli}{\al\over(\al\mli-1)}\bigg[
\nonumber\\
&&\left({m_c^2\al\mli\over2}-{2(M^2+\al{\mmi}^4)\over\al\mli-1}\right)\left(
{m_c^2\al^2\mli(M^2-\al{\mmi}^4+2\mli)\over M^2(\al\mli-1)^2}+{t\mli(\al M^2
+1)\over(M^2+\mli)^2}\right)
\nonumber\\
&+&{\al\mli\over2(\al\mli-1)}\left(
{8(M^2+\mli)\over \al\mli-1}-\al m_c^2\mli\right)\bigg],
\eeqa

\beqa
&&\Pi_G^{(d2)}(t,M^2,\mli)={{\mmi}^4\over12 M^2}\int_{1/\mli}^\infty
~d\al ~e^{-{m_c^2\al(M^2+\al{\mmi}^4)\over M^2(\al\mli-1)}}
~e^{t(\al\mli-1)\over M^2+\mli}{\al^2\over(\al\mli-1)^2}\bigg[
\nonumber\\
&&\left({m_c^2\al^2\mli(M^2+\mli)(M^2+2\mli-\al{\mmi}^4)\over M^2
(\al\mli-1)^2}+{t\mli(\al M^2+1)\over(M^2+\mli)}+\right.
\nonumber\\
&-&\left.{(M^2+2\mli+\al{\mmi}^2M^2)\over \al\mli-1}\right)\left(
{m_c^2\al^2{\mmi}^2(M^2+2\mli-\al{\mmi}^4)\over M^2(\al\mli-1)^2}+{t
\mli(\al M^2+1)\over(M^2+\mli)^2}\right)+
\nonumber\\
&+&
{m_c^2\al^2{\mmi}^2(M^2+\mli)(\al{\mmi}^6-\al{\mmi}^4-3\al\mli M^2-4\mli-M^2)
\over M^2(\al\mli-1)^3}+{m_c^2\al^2{\mmi}^4
\over M^2}\bigg(
\nonumber\\
&&\left.
{(M^2-\al{\mmi}^4+2\mli)\over(\al\mli-1)^2}\right)
+{tM^2\mli(\al M^2+1)\over(M^2+\mli)^2}
-{2t\al {\mmi}^4(\al M^2+1)\over(M^2+\mli)(\al\mli-1)}+
\nonumber\\
&+&
\left.{\mli((\al M^2+2)
(2\al\mli+1)+3\al M^2)\over(\al\mli-1)^2}
\right],
\eeqa

\beqa
&&\Pi_G^{(d3)}(t,M^2,\mli)=-{\mli\over3}\int_{1/\mli}^\infty
~d\al ~e^{-{m_c^2\al(M^2+\al{\mmi}^4)\over M^2(\al\mli-1)}}
~e^{t(\al\mli-1)\over M^2+\mli}{\al\over(\al\mli-1)}\bigg[
\nonumber\\
&-&\left.1+{m_c^2\al^2{\mmi}^4\over M^2
(\al\mli-1)}-{tM^2(\al\mli-1)\over(M^2+\mli)^2}\right],
\eeqa

\beqa
&&\Pi_G^{(d4)}(t,M^2,\mli)={1\over6 M^2}\int_{1/\mli}^\infty
~d\al ~e^{-{m_c^2\al(M^2+\al{\mmi}^4)\over M^2(\al\mli-1)}}
~e^{t(\al\mli-1)\over M^2+\mli}{1\over(\al\mli-1)}\bigg[
\nonumber\\
&&-{m_c^4\al^2(M^2+\al{\mmi}^4)\over 2(\al\mli-1)}
+{t\al\mli(3M^2+\mli-\al M^2\mli+\al{\mmi}^4)\over
(M^2+\mli)}-{m_c^2t\al^2{\mmi}^2\over 2}+
\nonumber\\
&+&{1\over(M^2+\mli)(\al\mli-1)}\bigg(
M^4(
-6+14\al\mli-9\al^2{\mmi}^4)+M^2\mli(-2+3\al\mli+
\nonumber\\
&-&3\al^2{\mmi}^4)+{\mmi}^4(
-1-\al\mli+\al^2{\mmi}^4)+
+m_c^2\al({\mmi}^4(1+4\al\mli-\al^2{\mmi}^4)+
\nonumber\\
&+&
2M^4(-1+3\al\mli)+
M^2\mli(1+6\al\mli+\al^2{\mmi}^4))\bigg)\bigg],
\eeqa

\beq
\Pi_G^{(d)}(t,M^2,\mli)=\Pi_G^{(d1)}+\Pi_G^{(d2)}+\Pi_G^{(d3)}+\Pi_G^{(d4)}
\eeq

\beq
\Pi_G^{(e)}(t,M^2,\mli)=-{{\mmi}^2\over 6M^2(M^2+\mli)}\int_{1/\mli}^\infty
~d\al ~e^{-{m_c^2\al(M^2+\al{\mmi}^4)\over M^2(\al\mli-1)}}
~e^{t(\al\mli-1)\over M^2+\mli}{M^2+\al{\mmi}^4\over\al\mli-1},
\eeq

\beqa
&&\Pi_G^{(f)}(t,M^2,\mli)={\mli\over6 M^2}\int_{1/\mli}^\infty
~d\al ~e^{-{m_c^2\al(M^2+\al{\mmi}^4)\over M^2(\al\mli-1)}}
~e^{t(\al\mli-1)\over M^2+\mli}{\al\over(\al\mli-1)^2}\bigg[
\nonumber\\
&&{m_c^2\al(M^2+\mli)^2\over M^2(\al\mli-1)}-2M^2-\mli-
{m_c^2\al^2{\mmi}^4(M^2+\mli)\over M^2(\al\mli-1)}+{tM^2(\al\mli-1)\over
(M^2+\mli)}+
\nonumber\\
&-&
-{m_c^2\al^2{\mmi}^2(M^2+\mli)(M^2+2\mli-\al{\mmi}^4)\over M^2(\al\mli-1)^2}
-{t\mli(\al M^2+1)\over(M^2+\mli)}+
\nonumber\\
&+&\left.
{2\al M^2\mli+\mli+\al{\mmi}^4\over\al\mli-1}\right],
\eeqa

\end{document}